# Reporting existing datasets for automatic epilepsy diagnosis and seizure detection

Palak Handa · Sakshi Tiwari · Nidhi Goel

**Abstract** More than 50 million individuals are affected by epilepsy, a chronic neurological disorder characterized by unprovoked, recurring seizures and psychological symptoms. Researchers are working to automatically detect or predict epileptic episodes through Electroencephalography (EEG) signal analysis, and machine, and deep learning methods. Good quality, open-source, and free EEG data acts as a catalyst in this ongoing battle to manage this disease. This article presents 40+ publicly available EEG datasets for adult and pediatric human populations from 2001-2023. A comparative analysis and discussion on open and private EEG datasets have been done based on objective parameters in this domain. Bonn and CHB-MIT remain the benchmark datasets used for the automatic detection of epileptic and seizure EEG signals. Meta-data has also been released for large EEG data like CHB-MIT. This article will be updated every year to report the progress and changing trends in the development of EEG datasets in this field.

Palak Handa
Dept. of ECE, DTU, Delhi
E-mail: palakhanda97@gmail.com

Sakshi Tiwari
Dept. of ECE, IGDTUW, Delhi
E-mail: sakshitiwari2911@gmail.com

Nidhi Goel
Dept. of ECE, IGDTUW, Delhi
E-mail: nidhi.iitr1@gmail.com

# 1 Introduction

Epilepsy is a chronic neurological disorder characterized by unprovoked, recurring (similar or different type) seizures. A clinical Electroencephalography (EEG) setting is used by doctors to observe different types of epileptic activity as it leaves distinct impressions in the form of inter-ictal epileptiform discharges, peri-ictal activities, and high-frequency oscillations *etc* [1].

The annual economic burden of epilepsy is enormous in developing countries like India where it is estimated to be 88.2% of Gross National Product (GNP) per capita, and 0.5% of the overall GNP [1]. Hence, early diagnosis through recent technologies like Artificial Intelligence (AI), feature engineering, data analytics, and multimedia is vital. It can aid in the Quality Of Life (QOF) of patients, and their associated caretakers.

Secured, reproducible AI algorithms, good quality data, and efficient computing horse power are the major elements for the development of early detection, and prediction of epileptic waveforms through EEG signals [2, 3]. There are several types of EEGs such as intracranial, scalp, ambulatory, *etc* through which good quality data can be generated. They are recorded in video, image, and signal format depending on their use and application in hospitals. There is a huge demand for real-time biomedical multimedia tools for data analysis, and pattern recognition of such formats. Bonn EEG time series database [4] was the first EEG dataset to be publicly available for research applications in this field. After Bonn data, 40+ datasets have been used in this domain. The datasets discussed in this paper encourage scientific advances in this field.

The data quality of biomedical datasets is measured through various factors such as the presence of artefacts and noise, missing values, descriptive information, an-notations by health experts, pre-fined data structure, processing and robustness to outliers in AI pipelines *etc* [5, 6, 7, 8, 9]. SOTA deep learning is being explored to correlate these parameters in biomedical datasets [7, 10, 11, 12, 13].

The datasets mentioned in Section 2 are freely available (except temple university which requires a login), re-distributable for research purposes, previously freely available and or became private or removed from the portal based on the adult and paediatric population. Table 1 shows a comparison of publicly available, open-access and private human and dog EEG datasets for an automated epilepsy diagnosis and seizure detection.

The source and availability of these were verified on 21-06-2023, which may change in the future. They were found using different keywords like 'EEG datasets for epilepsy', 'datasets for epilepsy detection', 'EEG based epilepsy diagnosis', and 'open EEG datasets' on popular databases such as Pub-med, Science Direct and google scholar.

## 2 EEG datasets for epilepsy diagnosis

There are several EEG datasets for epilepsy diagnosis that are freely available and private due to various reasons such as lack of ethical clearance, patient confidentiality, and project funding. This section lists all the existing EEG datasets with their URLs for the adult and pediatric populations in Section 2.1 and 2.2. Miscellaneous datasets have been mentioned in Section 2.3 and 2.4.

2.1 Adult datasets

The adult population consists of affected individuals above 20 years of age. There are several databases like the American Epilepsy Society Seizure Prediction Challenge database [14], dataset of EEG recordings of pediatric patients with epilepsy based on the 10-20 system [15] and Karunya University [16] which contain both adult and pediatric EEG data. They are mentioned in Section 2.2.

*2.1.1 Bonn EEG time series database [4]*

This database comprises 100 single channels EEG of 23.6 seconds with a sampling rate of 173.61 Hz. Its spectral bandwidth range is between 0.5 Hz and 85 Hz. It was taken from a 128-channel acquisition system. Five patients' EEG sets were cut out from a multi-channel EEG recording and named A, B, C, D and E. Sets A and B are the surface EEG recorded during the eyes closed and open situations of healthy patients respectively. Set C and D are the intracranial EEG recorded during a seizure-free from within the seizure-generating area and outside seizure generating area of epileptic patients respectively. Set E is the intracranial EEG of an epileptic patient during epileptic seizures. Each set contains 100 text files wherein each text file has 4097 samples of 1 EEG time series in ASCII code. A bandpass filter with cut-off frequencies of 0.53 Hz and 40 Hz has been applied to the data. It is artifact-free data and hence no prior pre-processing is required for the classification of healthy (non-epileptic) and unhealthy (epileptic) signals. The strong eye movement artifacts were omitted. It was made available in 2001. The extended version of this data is now a part of the EPILEPSIA project. Available link: a and b.

*2.1.2 Bern-Barcelona EEG database [17]*

This multi-channel EEG database was recorded using specialized electrodes and consists of five patients with longstanding pharmacoresistant temporal lobe epilepsy. The patients underwent epilepsy surgery. The sampling rate was either 512 or 1024 Hz based on whether they were recorded with less or more than 64 channels of the EEG system. Three out of five attained complete seizure freedom. Two types of EEG are present in the data *i.e.,* focal and non-focal. Each file has about 10240 samples for a time duration of 20 seconds. Available link: c and d

*2.1.3 Widespread EEG Changes Precede Focal Seizures [18]*

Low-pass filtered at 500 Hz and recorded at 2,000 Hz, depth electrode ictal EEG data were collected from 40 consecutive individuals with pharmacoresistant lesional focal epilepsy. The first EEG sign suggestive of seizure activity was chosen 30 seconds before and immediately before predefined EEG portions (immediate pre-ictal). (baseline). Standard (delta, theta, alpha, beta, and gamma) and high-frequency bands were discovered using discrete wavelet transform, spectral analysis, and visual examination. (ripples and fast ripples). Available link.

*2.1.4 Temple University EEG corpus [19]*

Temple University EEG corpus is the largest free EEG data available for epilepsy and seizure types diagnosis to date. It consists of data acquired from 2000 to 2013 using different EEG clinical settings for about 10,874 patients. Some of the EDF files contain EEG channels along with EKG, EMG, and photo stimuli recordings as well. The majority of the EDF files contain about 31 EEG channels. However, some EDF files also contain 20 EEG channels. The sampling frequency used to record the EEG data also varies and was found to be 250, 256, and 512 Hz. All the recordings are multi-channel in nature. The electrodes have been placed according to the 10-20 international EEG system. Such variations help can help in evaluating the reasons and dependence of these parameters for future generalized machine learning pipelines for an epilepsy diagnosis. This community has developed various software products such as annotation tools, toolboxes for seizure detection, and an EDF browser for data analysis of EEG, EMG, and ECG *etc* signals. EDF browser helps to view EEG recordings in a video form. There are about eight datasets available such as IBM Features For Seizure Detection (IBMFT), the TUH EEG epilepsy corpus, seizure corpus, slowing corpus, and events corpus *etc*. Various software, documentation files, annotations, and clinical and research grade information has been provided on their website. A user ID and password are required to get access to these datasets. Available link.

*2.1.5 Neurology and sleep center, New Delhi EEG dataset [20]*

This database comprises of 5.12 seconds EEG data. It was recorded using 57 EEG channel Grass Tele-factor Comet AS40 Amplification System; sampled at 200 Hz. Its spectral bandwidth range is between 0.5 Hz and 70 Hz. Time series EEG datasets are categorized into three major MATLAB file folders namely ictal, pre-ictal, and inter-ictal stages. Each MAT file has 1024 samples. A subset of this database is publicly available. Available link.

*2.1.6 Analysis of Epileptic Discharges from Implanted Subdural Electrodes in Patients with Sturge-Weber Syndrome [21]*

The majority of people with Sturge-Weber syndrome (SWS) suffer epilepsy, and surgery is required for half of them. This study included five intractable epileptic patients with SWS who had sub-dural electrodes inserted for pre-surgical assessment. Additionally, at SOZ, the EEG Complex Demodulation Method was used to do power spectrogram analysis on specific frequency bands beginning 60 seconds before the visually detected seizure began. (CDM). Five patients provided 21 seizures for assessment. The typical seizure duration and propagation speed were 19.4 33.6 min and 3.1 3.6 cm/min, respectively. At the SOZ, significant power spectrogram shifts at frequencies of 10–30 Hz and 30–80 Hz from 5–15 seconds before the start of a seizure were found. Available link.

*2.1.7 Epileptic Seizure Recognition Dataset [4]*

The time series EEG dataset consists of 11500 instances of EEGs of 4 subjects suffering from epilepsy. This data has been removed from the UCI machine learning repository recently and was released in 2017. It is a simplified version of the original data released by [4]. It consists of 5 subjects (4 unhealthy and 1 healthy) performing different activities and experiencing epileptic seizures except for subject 1. The time duration for each EEG was 23.5 seconds.

*2.1.8 Patient-specific seizure prediction based on heart rate variability and recurrence quantification analysis [22]*

Electrocardiogram and electroencephalogram signals in EDF format were collected from 15 patients with epilepsy. All of the patients underwent long-term video-EEG monitoring, ECG monitoring, and electrode placement based on the international 10-20 system. ECG measurements were made concurrently at a 512 Hz sampling rate. Available link.

*2.1.9 Data from: SYNGAP1 encephalopathy: a distinctive generalized developmental and epileptic encephalopathy [23]*

Patients were included who had chromosomes or SYNGAP1 mutations that were (presumably) pathogenic. employing a standardized epilepsy questionnaire, medical records, an EEG, an MRI, and seizure videos to analyze the phenotypes of individuals. SYNGAP1 mutations (n=53) or microdeletions (n=4) were found in 57 patients (53% men, median age 8 years). In 56/57 patients, epilepsy was present: generalized in 55, focal in 7, and infantile spasms in 1. The median age at which seizures began was 2. Available link.

*2.1.10 Dataset of simultaneous scalp EEG and intracranial EEG recordings and human medial temporal lobe units during a verbal working memory task [24]*

The nine subjects that participated in the verbal working memory task are represented in the dataset. Patients with epilepsy who were receiving intracranial monitoring for epileptic seizure localization served as the subjects. In a modified Sternberg task, the temporal separation of memory item encoding, maintenance, and recall was used. The dataset includes waveforms and spike timings of 1526 units recorded in the medial temporal lobe, intracranial EEG (iEEG) recorded with depth electrodes, concurrently acquired scalp EEG with the 10-20 system, and the MNI coordinates and anatomical describing of all intracranial electrodes. Available link.

*2.1.11 Intracranial EEG Epilepsy-Study 3 [25]*

The database is a record of seizures of a 21-year-old patient who has 11 seizures at the onset. This included 9 clinical seizures and two seizures that were sub-clinical. This data is, thus, a result of computer-assisted prolonged intracranial EEG monitoring sessions due to the presence of certain discharges. Available link.

*2.1.12 Intracranial EEG Epilepsy-Study 5 [26]*

The dataset is about a 16-year-old male patient, who is right-handed and was brought to the epilepsy monitoring unit for intracranial monitoring. A lengthy intracranial EEG monitoring session with computer assistance that lasted roughly two days is peculiar in this case. The patient had four complex partial seizures throughout the monitoring session. Pathological specimens of the cortex and underlying white matter show modifications that correspond to distant lethal injury. Available link.

*2.1.13 Intracranial EEG Epilepsy-Study 6 [27]*

This data has been obtained from an abnormal computer-assisted prolonged intracranial EEG monitoring session. The data was recorded from a 23-year-old female who underwent intracranial monitoring. The patient had 42 seizures, which were bilateral and non-localized. Available link.

*2.1.14 Single electrode EEG data of healthy and epileptic patients [28, 29]*

This dataset was generated with a motive to build a predictive epilepsy diagnosis model and publically available since 2020. It was generated on a similar acquisition and settings *i.e.,* sampling frequency, bandpass filtering, and a number of signals and time duration as of the University of Bonn. It has overcome the limitations faced by the University of Bonn dataset such as different EEG recordings (inter-cranial and scalp) for healthy and epileptic patients [28]. All the data were taken exclusively using surface EEG electrodes for 15 healthy and epileptic patients. Available link.

*2.1.15 Siena Scalp EEG Database [30, 31]*

This multi-channel EEG database of 14 epileptic patients (9 males and 5 females) was recorded using specialized amplifiers, and reusable electrodes. The signals were recorded with a sampling rate of 512 Hz and stored in EDF files. The data has been acquired from the Unit of Neurology and Neurophysiology of the University of Siena, Italy, and focuses on seizure prediction. It is an integral part of the national interdisciplinary research project PANACEE. This data contains 47 seizures from 128 hours of video EEG recording. The start and end time of a seizure was also recorded and contained the list of electrodes present on the scalp of a patient during event recording. Three types of seizures namely focal onset with and without impaired awareness, and focal to the bilateral tonic–clonic (FBTC) were found and recorded in the diseased patients. Available link.

*2.1.16 Electrographic Seizure detection by Neuro ICU nurses [32]*

This dataset is composed of digital EEG recordings that were obtained with electrodes placed according to International Placement System.cEEG recordings were processed through Persyst 12 to yield 1-hour qEEG trends.qEEG consists of a rhythmic spectrogram and amplitude-integrated EEG. The author of the study CBS determined the seizure's length, amplitude, and backdrop. The seizure's geographic extent (focal, hemispheric, or generalized/secondary generalized) was also determined. Available link.

*2.1.17 Epileptic EEG Dataset [33]*

This multi-channel, long-term EEG database was recorded for 6 patients suffering from focal epilepsy. They were undergoing pre-surgical evaluation for possible e-epilepsy surgery. Different EEG segments of a seizure like ictal, pre-ictal, inter-ictal, and its onsets have been included in the data. The signals were recorded with a sampling rate of 500 Hz and stored in EDF files. Labeled and classified data points (train and test set) have been mentioned for complex partial electrographic, and video-detected seizures. All the EEG signals underwent bandpass filtering of range 1-70 Hz where 50 Hz (utility frequency) was also removed. Available link.

*2.1.18 Ictal source imaging in epilepsy patients [34]*

For development and testing, the dataset consists of EEGs from 64 patients (37 hours of recordings with over 400 seizures) and 30 controls who are age and sex-matched (9 hours of recordings). The detector, which examined two bipolar EEG channels, had a sensitivity of 97.6% and a detection rate of 0.7/h for seizures lasting longer than two seconds. All erroneous detections in the patients were connected to epileptiform discharges. The false detection rate decreased to 0.5/h when the time threshold was increased to 3s. The overlap between the seizures that were automatically recognized and those that actually occurred was 96%. The one-channel processing performance for mid-range smartphones running Android 10 (approximately 0.2 s every 1 min of EEG) was fast enough for real-time seizure identification for EEG recordings recorded at 250 Hz. Available link.

### 2.1.19 Intracranial EEG dataset 1 [35]

This dataset includes data recorded from patients who had epilepsy while performing a multisource interference task. Some of them received electrical stimulation during MSIT trials. Available link.

### 2.1.20 EEG Dataset of Epileptic Seizure Patients [36]

This dataset comprises EEG data from 40 patients having epileptic attacks, who are 4 to 80 years of age (Thus, it comes under the Paediatric dataset too). The Allengers VIRGO EEG machine at Medisys Hospitals in Hyderabad, India, generated the raw data. The placement of the EEG electrodes mirrored the 10–20 international standard. 16 channels of EEG data were captured at 256 samples per second. Available link.

### 2.1.21 Pre-processed EEG dataset with epileptic seizure [37]

This dataset consists of multi-channel, good-quality EEG recordings of eleven patients where two of them suffered from seizures. The EEG signals were captured from 16 channels. It was observed that the seizure lasts for 170 seconds. These signals were recorded with a sampling rate of 256 Hz and were stored in a .csv format. Available link.

### 2.1.22 Dataset of rhythmicity spectrogram based images of seizure and non-seizure EEG signals [2]

The dataset includes 105 frames from chb01, 30 frames from chb02, 90 frames from chb05, and 75 frames from chb05, all of which were taken individually from seizure and ictal edf files. Twenty ictal signals and 600 total image frames make up the total for the ictal time, which is 25 minutes. Seizure and non-seizure EEG images have been placed in the form of png in the train, test, and validate folders of the dataset. For the objective of distinguishing between seizure- and non-seizure-related EEG pictures, it can be included in the machine and deep learning pipelines. Available link.

### 2.1.23 Epilepsy seizure prediction [38]

The dataset from the reference consisted of 5 folders with 100 files each and each file represents a single subject. The files contain brain activity for 23.6 seconds each. They have been sampled into 4097 data points and were recorded from 500 patients. The data points have been shuffled into 23 chunks with each chunk containing 178 data points for 1 second. So, there are 11500 rows with 178 data points for 1 second(columns.) Available link.

### 2.1.24 Epileptic High Frequency Oscillation EEG with Network Analysis [39]

Data from two patients from the University of Michigan are included in the dataset, which shows the functional linkage of HFOs throughout all intracranial channels. The functional connectivity analysis of medical intracranial EEG recordings from patients who underwent reconstructive surgery has been used to describe HFO networks. Available link.

### 2.1.25 Simultaneous EEG-SEEG recordings of deep mesial temporal sources [40]

Seven patients (three females, mean age 38 years) with temporal lobe epilepsy were incorporated in 9 simultaneous EEG-SEEG datasets matching mesial temporal networks. (TLE). An interictal intracerebral spike (IIS) source and an epileptogenic zone restricted to the temporal lobe were present in these patients. 1,949 IIS in total were chosen. IIS amplitude was 729 279 V on average. The identical 1 sec epoch that is centred on t0 was used to extract EEG-SEEG segments. Available link.

### 2.1.26 Epilepsy-iEEG-Multi center dataset [15]

The epilepsy-iEEG-Multi centre dataset contains EEG and i-EEG of 100 subjects from five medical centres such as NIH, JHH, UMMC, and UMF. It was prepared for release as part of work done by Bernabei and Li et al. This dataset contains one seizure per EDF file. Within each EDF file, there is an event marker that has been annotated by experienced clinicians. This marking has been done electrographically via approaches in epilepsy clinical workflow at the five medical centres. The files also contain information about the occurrence and onset situation of the annotated seizures. Available link.

*2.1.27 Annotated epileptic EEG components for artifact reduction [41]*

This database is composed of a source of annotated electroencephalogram-independent components acquired from epileptic patients. This dataset is composed of 77,246 components obtained from 613 hours of electroencephalogram, which is a non-invasive multichannel bio-signal that records the brain's electrical activity. Available link.

2.2 Pediatric datasets

The paediatric EEG database consists of affected individuals from age 1 month-20 years. Australian EEG [42] and Karunya University [16] data contains both adult and paediatric patient data.

*2.2.1 The Australian EEG Database [42]*

This database contains about 18,500 EEG recordings of diseased patients aged between 24 months to 90 years. Preliminary results achieved from the two conducted studies have been discussed through this data. It was developed in collaboration between staff from the University of Newcastle and the John Hunter Hospital (JHH) wherein 11 years of EEG data has been extracted from different patients, neurological diseases, gender, and their clinical inferences and reports. About 859 Grand mals or GTC seizures are present in this database. It was previously available at link. Now, it can be accessed upon request from the corresponding authors at aed-@newcastle.edu.au [43]. The processed version of Australian EEG data in BVA format is available at (link).

*2.2.2 Children's hospital Boston–MIT database [44]*

This database comprises 844 hours of continuous EEG. 23 pediatric patients from age 1.5-19 who underwent scalp multi-channel EEG recording. It is the first paediatric EEG database available for epilepsy and seizure diagnosis. The patients were given anti-seizure medications. About 200 seizures were recorded in a universal bio-polar montage with about 24-27 EEG channels. The sampling frequency was kept to 256 Hz. Each EEG segment is called a record which usually is for a duration of one hour. There are 9-42 edf files from a single subject. Additional vagal nerve stimulus signals are also present. Separate file names and montages have been mentioned for seizure v/s non-seizure episodes in EEG segments. Available link.

*2.2.3 Karunya University [16]*

This database comprises 18 channels of EEG data with segments of normal, focal, and generalized epileptic seizure activities from 1–107 years of patients. It was released in 2014 but the website is not available for research use now. Each segment has 2056 sample points. The sampling frequency was kept to 256 Hz. The EEG recordings vary from 40 minutes to one hour. It was collected from a diagnostic center based in Coimbatore, India. Available link.

*2.2.4 A dataset of neonatal EEG recordings with seizures annotations [45]*

This database consists of multi-channel, good-quality EEG recordings of 79 term neonates where 39 of them suffered from neonatal seizures in the NICU of Helsinki University Hospital, Finland. The recordings were captured with a NicOne EEG amplifier and 19 EEG channel caps. The signals were recorded with a sampling rate of 256 Hz and stored in EDF files. It consists of seizure annotations by healthcare experts for seizure detection purposes. The data was pre-processed using Butterworth high-pass filtering. The data also contains natural artifacts. Available link.

*2.2.5 Dataset of EEG recordings of pediatric patients with epilepsy based on the 10-20 system [15]*

This dataset consists of scalp EEG recordings to study the impact of age on observed High-Frequency Oscillations (HFO) in pediatric epileptic patients. Three hours of pediatric and adult EEG sleep data were recorded for 30 focal or generalised epileptic patients. The signals were recorded with a sampling rate of 2000 Hz and stored in EDF files. Different sleep stage annotations are available in this database. Available link.

2.3 EEG Meta-data

The concept of meta-data for large EEG datasets has been proposed recently [46, 47]. Here, 'Meta-data' refers to a sub-set of large EEG database *i.e.,* the available data is a processed form of original data in a certain scientific manner.

*2.3.1 Pre-processed CHB-MIT Scalp EEG database [46]*

This database is a pre-processed version of CHB-MIT Scalp EEG data and has been released to promote balanced data development in epileptic seizure detection and prediction using machine and deep learning methods. It was released in 2021 on IEEE data port website. The data contains 4096 seconds (68 minutes) of pre-ictal and ictal EEG for all 24 patients. Five different comma-separated value (csv) files are released. The first two files contain raw ictal and pre-ictal EEGs with varied EEG channels. The next two files contain ictal and pre-ictal EEGs with 23 EEG channels. The last file contains balanced data of pre-ictal and ictal EEG segments with 23 channels. The class labels have been inserted in the last column of this csv file. Further data processing and artifact removal have not been done. It is about 5.12 GBs which is about 8 times less than the original data space. Available link.

*2.3.2 Meta-EEG of CHB-MIT Scalp EEG database [47]*

This database consists of balanced, fixed time and length, peri-ictal and non-seizure multi-channel EEG fragments extracted from the original CHB-MIT scalp EEG data. The need of developing a seizure-sensitive, meta-EEG for machine learning applications in large EEG databases like CHB-MIT has been elucidated through this data. The peri-ictal EEGs constituted of 10 seconds of pre-ictal and post-ictal activity with the full length of ictal activity as mentioned in the annotation file of the original data. Inter-ictal states are not included in this data. Presently, the data is not open-sourced and requires permission from the corresponding author.

2.4 Others

This section discusses EEG datasets with partial information on the patient's age and no. of channels *etc.,* and contains animal data along with human EEG data.

*2.4.1 American Epilepsy Society Seizure Prediction Challenge database [14]*

This database consists of intra-cranial EEG segments from dogs and humans with different acquisitions of sampling rate, duration of EEGs, and no. of electrodes *etc.* It was released as a part of the Kaggle challenge hosted by the American Epilepsy Society in 2014 for the development of seizure forecasting systems and witnessed about 504 teams. Different seizure segments of ictal, pre-ictal, post-ictal, and inter-ictal were provided in MATLAB files. The data storage was about 105 GB. Additional annotated EEG data was also provided by the University of Pennsylvania and the Mayo Clinic. Available link.

*2.4.2 Epilepsy Ecosystem Seizure Prediction data [48]*

This dataset contains long-term 16-channel, intra-cranial EEG recordings from the world-first clinical trial of the implantable NeuroVista Seizure Advisory System, recorded at St Vincent's Hospital in Melbourne, Australia. It was released after the success of 'The melbourne-University AES-MathWorks-NIH Seizure Prediction Challenge' challenge hosted on Kaggle in 2016. It contains more than 1362 seizure episodes obtained from three female patients. The dataset can be accessed upon registering on the website. Available link.

*2.4.3 Spike-wave seizures and p̈roepilepticäctivity in WAG/Rij rat model of absence epilepsy as recorded in the cortex, hippocampus and thalamus [49]*

A male WAG/Rij rat aged 8 months had his behavioral sleep recorded. Bipolar recordings: Channel 2 (green) represents the ventrobasal thalamus, Po; Channel 1 (red) represents the hippocampus, CA1 (locally). (locally). Monopolar recordings with the cerebellum serving as the reference The frontal cortex is represented by channel 3 in blue, the hippocampus, CA1, by channel 4 in green, and the ventrobasal thalamus, Po, by channel 5. Spike-wave discharges from absence seizures are denoted as "SWD," and proepileptic activity episodes are denoted as "ProEpi." "Proepileptic" activity manifests as high voltage spindles during behavioural sleep or as primitive SWD occurring before epileptic SWD. Available link.

*2.4.4 Dataset of EEG power integral in soman-exposed rats [50]*

The database is determined to find the efficacy of doses of midazolam therapy when administered at a delayed time point in protection against induced seizure severity. The EEG recordings were analyzed with MATLAB algorithm and quantitative measures of seizure onset and duration, and also their spontaneous recurrence. The folder contains raw EEG data which has been compressed into zip format into 3-4 volumes. This might help in seizure detection and protection against neuroprotective for soman-epilepticus. Available link.

*2.4.5 Canine Epilepsy Dataset [14]*

Three dogs with naturally occurring focal epilepsy that was later generalised are included in the dataset. The cause of their epilepsy is uncertain. These canines were implanted with four bilateral 4-contact subdural strip electrodes, providing them with eight connections in each hemisphere. The intracranial EEGs of these dogs were continuously recorded throughout periods of 45.8, 451.9, and 475.7 days, respectively. In every dog, the seizure symptoms were the same. Dogs 2 and 3 began receiving phenobarbital on days 105 and 54, respectively, following the death of dog 1 from status epilepticus. As a result, there were no further episodes of status epilepticus and seizures were substantially less frequent. Available link.

### 2.4.6 Multi-center intracranial EEG dataset [51]

This dataset contains 155,182 and 193,118 intracranial EEG data clip recordings of 39 epileptic patients from Mayo Clinic, MN, USA and St. Anne's University Hospital Brno, Czech Republic respectively. It was released in 2020 to facilitate research in generalized machine learning and advanced signal processing methods for the analysis of intracranial EEG recordings. The data has been subdivided into three class labels namely physiological activity, pathological/epileptic activity, and artifact containing signals. The EEG signals are further supported by clinically useful features and annotations such as information on seizure onset zone, anatomical location for seizure localization, noise/artifact time stamps, and electrode placing *etc.* The annotated EEG signals are 3 seconds in duration. The data and annotations are stored in .csv and .mef file format respectively with total data storage of up to 24.3 GBs. Python and MATLAB code files are also attached. Available link.

### 2.4.7 Epileptic seizure detection using machine learning techniques [52]

The diagnosis system that was created using machine learning techniques is included in this dataset, along with the raw data (Dataset folder), processed dataset (newfeature.csv; after feature extraction), and Python code (code.py). utilise the code.py file, which has the different classifier performance commented statements, to utilise different classifiers for classification. To obtain the classification result, one can uncomment the statement linked to the classifier you want to employ. Available link.

### 2.4.8 EEG of Genetic Absence Epilepsy Rats (GAERS) [53]

This database contains spike and wave discharge of rats suffering from epilepsy. The data is split into data files with a recording length of up to ten minutes. There are two classes in the dataset wherein a 0 class label indicates to no seizure and 1 indicates seizure activity.

The data also contains additional images and information about non-physiological signals. A neural network-based pipeline has been proposed using this dataset called Genetic Absence Epilepsy Rats (GAERS). Available link.

### 2.5 Private databases

Several private databases have also been recorded for epilepsy diagnosis using EEG signals [28, 54, 55, 56, 57, 58, 59, 60, 61, 62]. The European Epilepsy database [63] is a private database which consists of high-quality, annotated EEG signals from the University of Bonn [4], Freiburg [64], Flint hills, and many multi-modal like MRI imaging data. The website of [16] is not available. EEG data in [65] was freely available till 2015 when its portal crashed.

## 3 Conclusion and future directions

Diagnosis, treatment, and management of epilepsy is still a challenging task for the scientific and healthcare community. Its detection by visual introspection of long-hour EEG is not only time taking but a very tedious and subjective task. AI can help in escalating this process and lead to the successful detection of different types of epilepsies through efficient, high-quality, and annotated EEG data.

This article has presented all the existing EEG datasets for epilepsy diagnosis with its availability, recent experimental AI pipelines, and comparative analysis between them. Reported EEG datasets have been collected using different clinical settings, motivation, research questions, and patient information. A rise in open-source, raw EEG datasets has been observed since 2019. With changing trends in AI, the introduction of transfer learning, meta-learning, and deep learning methods, the handling of bigger EEG datasets like temple university corpus has become possible. Increasing data leads to higher computational power and processing and thus, researchers have also proposed specific EEG channel data analysis through channel selection and other AI methods, meta-data-based machine learning analysis in comparison to using continuous and raw EEG data. A generalized data development pipeline is a way forward to universal AI pipelines in automated epilepsy diagnosis.

Interdisciplinary collaborations of biomedical engineers, machine learning engineers, and medical specialists can help in annotating the existing datasets into different types of seizures, the presence of artifacts and noises, the inclusion of important clinical information in AI pipelines, and its dependence on detection and prediction pipelines. Such datasets not only motivate scientific research for AI-based real-time early diagnosis of epilepsy but also promotes awareness amongst the machine learning, medical and general community to combat this deadly disease and somehow aid in improving the QOF of patients suffering from epilepsy and seizure disorder.

**Table 1** Comparison of existing EEG datasets for an epilepsy diagnosis.

| Ref. | Availability | Type | Source | Year | Size | No. of channels | No. of patients | Sampling frequency | EEG segments |
|---|---|---|---|---|---|---|---|---|---|
| [4] | Freely available | Adult | e-repositori upf. | 2001 | 3.05 MB | 100 single | 5 | 173.61 Hz | seizure states, healthy |
| [42] | Upon request | Paediatric and adult | - | 2005 | - | - | - | - | - |
| [44] | Freely available | Paediatric | PhysioNet repository | 2010 | 42.6 GB | 23-27 | 23 | 256 Hz | Intractable seizures |
| [17] | Freely available | Adult | e-repositori upf. | 2012 | 814 MB | 64 | 5 | 512 Hz | Focal, Non-focal |
| [18] | Freely available | Adult | Figshare | 2013 | 5.61 MB | - | 40 | 2000 Hz | Ictal, preictal, seizure onset |
| [14] | Freely available | Dog and human | Kaggle | 2014 | 105 GB | - | - | - | different types |
| [16] | Not available | Paediatric and adult | Website | 2014 | - | - | - | 256 Hz | normal, focal and generalized epileptic seizures |
| [19] | Free but requires login | Adult | Website | 2015 | 572 GB | 20-31 | 10,874 | 250, 256, 512 Hz | different types |
| [20] | Freely available | Adult | Researchgate | 2016 | 604 KB | 57 | 10 | 200 Hz | Ictal, inter-ictal, pre-ictal EEGs |
| [21] | Freely available | Adult | Figshare | 2016 | 120.78 MB | - | 5 | 30-80 Hz | Seizure onset using EEG CDM |
| [4] | Removed | Adult | UCI repository | 2017 | 3 MB | 100 single | 5 | 173.61 Hz | seizure related, healthy |
| [45] | Freely available | Paediatric (neonates) | Zenedo | 2018 | 4.3 GB | 19 | 79 | 256 Hz | seizure onset |
| [48] | Requires registration | Adult | Website | 2018 | - | 16 | 3 | 400 Hz | Seizure episodes |
| [22] | Freely available | Adult | Zenodo | 2018 | 14.5 GB | - | 15 | 512 Hz | Seizure, epilepsy |
| [49] | Freely available | Rat | Mendeley Data | 2018 | 36 MB | 5 | 1 | 1 kHz | Rat Model, Video-EEG, Absence Seizures |
| [55] | Private | Adult | - | 2019 | - | 19 | 115 | 128 Hz | epileptic and healthy |
| [28] | Private | Adult | - | 2019 | - | - | 50 | 250, 256 Hz | generalized and focal epilepsies |
| [56] | Private | Adult | - | 2019 | - | 21 | 5 | 500 Hz | focal and tonic-clonic |
| [57] | Private | Pediatric | - | 2019 | - | - | 29 | 200, 500 Hz | typical absence seizures |
| [60] | Private | Adult | - | 2019 | - | - | 12 | 256 Hz | seizure events |
| [61] | Private | - | - | 2019 | - | 21 | 25 | 200 Hz | seizure events |
| [62] | Private | - | - | 2019 | - | 18 | 10 | 256 Hz | seizure states |
| [58] | Private | - | - | 2019 | - | 22 | 22 | 250 Hz | ictal, non-ictal |
| [23] | Freely available | Adult | Zenodo | 2019 | 90.7 KB | - | 57 | - | Myoclonic, typical, atypical atonic seizures |
| [24] | Freely available | Adult | G-node | 2019 | 16 GB | - | - | 4-32 kHz | Intracranial EEG |
| [25] | Freely available | Adult | Pennsieve | 2019 | 33.79 GB | - | 1 | - | Intracranial monitoring |
| [26] | Freely available | Adult | Pennsieve | 2019 | 8.61 GB | - | 1 | - | Intracranial monitoring |
| [27] | Freely available | Adult | Pennsieve | 2019 | 34.88 GB | - | 1 | - | Intracranial monitoring |
| [50] | Freely available | Rats | Mendeley Data | 2019 | 7223 MB | - | - | - | Soman-induced seizure severity, epileptogenesis |
| [14] | Freely available | Dogs | Pennsieve | 2019 | 263.62 GB | - | 3 | - | Intracranial EEG, seizure symptomatology |
| [28, 29] | Freely available | Adult | Zenedo | 2020 | 20.3 MB | - | 15 | 173.61 Hz | Inter-ictal |
| [59] | Private | - | - | 2020 | - | 21 | - | 250 Hz | seizure onsets |
| [54] | Private | Adult | - | 2020 | - | 21 | 150 | 256 Hz | seizure and normal |
| [30, 31] | Freely available | Adult | PhysioNet repository | 2020 | 20 GB | 29 | 14 | 512 Hz | Epileptic seizures (focal onset, tonic-clonic) |
| [51] | Freely available | - | Figshare | 2020 | 24.3 GBs | - | 39 | - | divided based on activity |
| [32] | Freely available | Adult | Repository.duke | 2020 | 43.8 KB | - | - | 128 Hz | Brief rhythmic discharges, periodic |
| [52] | Freely available | - | Mendeley Data | 2020 | 2 MB | - | - | - | Epilepsy dataset |
| [33] | Freely available | Adult | Mendeley repository | 2021 | 3133 MB | 21 | 6 | 500 Hz | Complex partial, electrographic and video-detected seizures |
| [15] | Freely available | Paediatric and adult | Open neuro repository | 2021 | 15 GB | 52 | 30 | 2000 Hz | HFO markings |
| [46] | Freely available | Paediatric | IEEE dataPort | 2021 | 5.12 GBs | 23-96 | 24 | 256 Hz | Ictal and pre-ictal EEGs |
| [47] | Private | Paediatric | - | 2021 | - | 22 | 23 | 256 Hz | peri-ictal and non-seizure EEGs |
| [34] | Freely available | Adult | Dryad | 2021 | 39.6 MB | - | 39 | - | Interictal epileptic discharges, seizure onset zone |
| [35] | Freely available | Adult | Zenodo | 2021 | 49.1 GB | - | - | - | Intractable epilepsy |
| [36] | Freely available | Adult | IEEE-dataport | 2021 | 153.77 MB | 16 | 40 | 256 Hz | Epileptic seizure |
| [53] | Freely available | Rats | IEEE-dataport | 2021 | 14.43 GB | - | - | - | Genetic epilepsy |
| [37] | Freely available | Adult | IEEE dataPort | 2022 | 15 MB | 16 | 11 | 256 Hz | Epileptic Seizures discharges, rhythmic delta activity |
| [2] | Freely available | Adult | Zenodo | 2022 | 49.9 MB | 15 | 16 | 256 Hz | Channel-wise ictal, random non-seizure |
| [38] | Freely available | Adult | Paperswithcode | 2022 | - | - | 500 | 0.04 Hz | Epileptic seizure |
| [39] | Freely available | Adult | Pennsieve | 2022 | 144.92 GB | - | - | 80-500 Hz | Intracranial EEG data |
| [40] | Freely available | Adult | Mendeley Data | 2022 | 347 MB | 128 | 7 | 512 Hz | Temporal lobe, interictal intracerebral spike (IIS) source confined to the mesial temporal |
| [15] | Freely available | Adult | Open Neuro | 2022 | 10.32 GB | - | 100 | - | ECoG and SEEG |
| [41] | Freely available | Adult | Zenodo | 2022 | 38.1 GB | - | - | - | - |

## References

1. SV Thomas, PS Sarma, M Alexander, L Pandit, L Shekhar, C Trivedi, and B Vengamma. Economic burden of epilepsy in india. *Epilepsia*, 42(8):1052–1060, 2001.
2. Palak Handa and Nidhi Goel. Epileptic seizure detection using rhythmicity spectrogram and cross-patient test set. In *2021 8th International Conference on Signal Processing and Integrated Networks (SPIN)*, pages 898–902. IEEE, 2021.
3. Palak Handa, Esha Gupta, Satya Muskan, and Nidhi Goel. A review on software and hardware developments in automatic epilepsy diagnosis using eeg datasets. *Expert Systems*, page e13374.
4. Ralph G Andrzejak, Klaus Lehnertz, Florian Mormann, Christoph Rieke, Peter David, and Christian E Elger. Indications of nonlinear deterministic and finite-dimensional structures in time series of brain electrical activity: Dependence on recording region and brain state. *Physical Review E*, 64(6):061907, 2001.
5. Susanta Kumar Rout, Mrutyunjaya Sahani, Chinmayee Dora, Pradyut Kumar Biswal, and Birendra Biswal. An efficient epileptic seizure classification system using empirical wavelet transform and multi-fuse reduced deep convolutional neural network with digital implementation. *Biomedical Signal Processing and Control*, 72:103281, 2022.
6. M Madhurshalini, Chitra Nair, and Nidhi Goel. Automatic identification of skin lesions using deep learning techniques. In *2020 IEEE/ITU International Conference on Artificial Intelligence for Good (AI4G)*, pages 230–235. IEEE, 2020.
7. Arti Anuragi, Dilip Singh Sisodia, and Ram Bilas Pachori. Epileptic-seizure classification using phase-space representation of fbse-ewt based eeg sub-band signals and ensemble learners. *Biomedical Signal Processing and Control*, 71:103138, 2022.
8. Li Rong Wang, Limsoon Wong, and Wilson Wen Bin Goh. How doppelgänger effects in biomedical data confound machine learning. *Drug discovery today*, 2021.
9. Vidhi Bishnoi, Nidhi Goel, and Akash Tayal. Wrapper-based best feature selection approach for lung cancer detection. In *International Conference on Artificial Intelligence and Sustainable Computing*, pages 175–186. Springer, 2021.
10. Samarjeet Kaur and Nidhi Goel. A dilated convolutional approach for inflammatory lesion detection using multi-scale input feature fusion (workshop paper). In *2020 IEEE Sixth International Conference on Multimedia Big Data (BigMM)*, pages 386–393. IEEE, 2020.
11. Erdem Tuncer and Emine Doğru Bolat. Classification of epileptic seizures from electroencephalogram (eeg) data using bidirectional short-term memory (bi-lstm) network architecture. *Biomedical Signal Processing and Control*, 73:103462, 2022.
12. Nidhi Goel, Samarjeet Kaur, Deepak Gunjan, and SJ Mahapatra. Dilated cnn for abnormality detection in wireless capsule endoscopy images. *Soft Computing*, pages 1–17, 2022.
13. Bin Gao, Jiazheng Zhou, Yuying Yang, Jinxin Chi, and Qi Yuan. Generative adversarial network and convolutional neural network-based eeg imbalanced classification model for seizure detection. *Biocybernetics and Biomedical Engineering*, 42(1):1–15, 2022.
14. J Jeffry Howbert, Edward E Patterson, S Matt Stead, Ben Brinkmann, Vincent Vasoli, Daniel Crepeau, Charles H Vite, Beverly Sturges, Vanessa Ruedebusch, Jaideep Mavoori, et al. Forecasting seizures in dogs with naturally occurring epilepsy. *PloS one*, 9(1):e81920, 2014.
15. Dorottya Cserpan et al. Dataset of eeg recordings of pediatric patients with epilepsy based on the 10-20 system, 2021.
16. Thomas George Selvaraj, Balakrishnan Ramasamy, Stanly Johnson Jeyaraj, and Easter Selvan Suviseshamuthu. Eeg database of seizure disorders for experts and application developers. *Clinical EEG and neuroscience*, 45(4):304–309, 2014.
17. Ralph G Andrzejak, Kaspar Schindler, and Christian Rummel. Nonrandomness, nonlinear dependence, and nonstationarity of electroencephalographic recordings from epilepsy patients. *Physical Review E*, 86(4):046206, 2012.
18. Piero Perucca, François Dubeau, and Jean Gotman. Widespread eeg changes precede focal seizures. *PloS one*, 8(11):e80972, 2013.
19. Iyad Obeid and Joseph Picone. The temple university hospital eeg data corpus. *Frontiers in neuroscience*, 10:196, 2016.
20. P Swami, B Panigrahi, S Nara, M Bhatia, and T Gandhi. Eeg epilepsy datasets, 2016.

21. Yasushi Iimura, Hidenori Sugano, Madoka Nakajima, Takuma Higo, Hiroharu Suzuki, Hajime Nakanishi, and Hajime Arai. Analysis of epileptic discharges from implanted subdural electrodes in patients with sturge-weber syndrome. *PLoS One*, 11(4):e0152992, 2016.
22. Lucia Billeci, Daniela Marino, Laura Insana, Giampaolo Vatti, and Maurizio Varanini. Patient-specific seizure prediction based on heart rate variability and recurrence quantification analysis. *PloS one*, 13(9):e0204339, 2018.
23. Danique RM Vlaskamp, Benjamin J Shaw, Rosemary Burgess, Davide Mei, Martino Montomoli, Han Xie, Candace T Myers, Mark F Bennett, Wenshu XiangWei, Danielle Williams, et al. Syngap1 encephalopathy: A distinctive generalized developmental and epileptic encephalopathy. *Neurology*, 92(2):e96–e107, 2019.
24. Ece Boran, Tommaso Fedele, Adrian Steiner, Peter Hilfiker, Lennart Stieglitz, Thomas Grunwald, and Johannes Sarnthein. Dataset of human medial temporal lobe neurons, scalp and intracranial eeg during a verbal working memory task. *Scientific data*, 7(1):30, 2020.
25. Intracranial eeg epilepsy - study 3 - blackfynn discover. https://discover.pennsieve.io/datasets/14, . (Accessed on 04/14/2023).
26. Benjamin H Brinkmann, Nicholas M Gregg, and Gregory A Worrell. Seizure forecasting in epilepsy: From computation to clinical practice. *Epilepsy*, pages 451–490, 2021.
27. Intracranial eeg epilepsy - study 6 - blackfynn discover. https://discover.pennsieve.io/datasets/15, . (Accessed on 04/14/2023).
28. Siddharth Panwar, Shiv Dutt Joshi, Anubha Gupta, and Puneet Agarwal. Automated epilepsy diagnosis using eeg with test set evaluation. *IEEE Transactions on Neural Systems and Rehabilitation Engineering*, 27(6):1106–1116, 2019.
29. Siddharth Panwar. Single electrode EEG data of healthy and epileptic patients, February 2020. URL https://doi.org/10.5281/zenodo.3684992.
30. Paolo Detti, Giampaolo Vatti, and Garazi Zabalo Manrique de Lara. Eeg synchronization analysis for seizure prediction: A study on data of noninvasive recordings. *Processes*, 8(7):846, 2020.
31. P. Detti. Siena scalp eeg database (version 1.0.0), 2020. URL https://doi.org/10.13026/5d4a-j060.
32. Safa Kaleem, Jennifer Kang, and Christa Swisher. Early electrographic seizure detection by neuro icu nurses via bedside real-time quantitative eeg (4542), 2020.
33. Wassim Nasreddine. Epileptic eeg dataset, 2021. URL https://doi.org/10.17632/5pc2j46cbc.1.
34. DS Eliashiv, SM Elsas, K Squires, I Fried, and Jr Engel. Ictal magnetic source imaging as a localizing tool in partial epilepsy. *Neurology*, 59(10):1600–1610, 2002.
35. Intracranial eeg dataset 1 — zenodo. https://zenodo.org/record/5083120#.ZDk4texBxQK, . (Accessed on 04/14/2023).
36. Eeg dataset of epileptic seizure patients — ieee dataport. https://ieee-dataport.org/documents/eeg-dataset-epileptic-seizure-patients. (Accessed on 04/14/2023).
37. Deepa B and Ramesh K. Preprocessed eeg dataset with epileptic seizure from snmc bagalkot, india. 2022. doi: 10.21227/q6y0-e695. URL https://dx.doi.org/10.21227/q6y0-e695.
38. Ahnaf Akif Rahman, Fahim Faisal, Mirza Muntasir Nishat, Muntequa Imtiaz Siraji, Lamim Ibtisam Khalid, Md Rezaul Hoque Khan, and Md Taslim Reza. Detection of epileptic seizure from eeg signal data by employing machine learning algorithms with hyperparameter optimization. In *2021 4th International Conference on Bio-Engineering for Smart Technologies (BioSMART)*, pages 1–4. IEEE, 2021.
39. Network dynamics of hfos - blackfynn discover. https://discover.pennsieve.io/datasets/280. (Accessed on 04/14/2023).
40. Laurent Koessler, Thierry Cecchin, Sophie Colnat-Coulbois, Jean-Pierre Vignal, Jacques Jonas, Hervé Vespignani, Georgia Ramantani, and Louis Georges Maillard. Catching the invisible: mesial temporal source contribution to simultaneous eeg and seeg recordings. *Brain topography*, 28:5–20, 2015.
41. Fábio Lopes, Adriana Leal, Júlio Medeiros, Mauro F Pinto, António Dourado, Matthias Dümpelmann, and César Teixeira. Epic: Annotated epileptic eeg independent components for artifact reduction. *Scientific Data*, 9(1):512, 2022.
42. MRLL Hunter, Robert LL Smith, Wendy Hyslop, Osvaldo A Rosso, R Gerlach, JAP Rostas, DB Williams, and Frans Henskens. The australian eeg database. *Clinical EEG and neuroscience*, 36(2):76–81, 2005.
43. A comprehensive list of asc data resources. URL https://autismresearchlab.group.shef.ac.uk/ASD_resources.html.
44. Ary L Goldberger, Luis AN Amaral, Leon Glass, Jeffrey M Hausdorff, Plamen Ch Ivanov, Roger G Mark, Joseph E Mietus, George B Moody, Chung-Kang Peng, and H Eugene Stanley. Physiobank, physiotoolkit, and physionet: components of a new research resource for complex physiologic signals. *circulation*, 101(23):e215–e220, 2000.
45. Nathan Stevenson, Karoliina Tapani, Leena Lauronen, and Sampsa Vanhatalo. A dataset of neonatal EEG recordings with seizures annotations, June 2018. URL https://doi.org/10.5281/zenodo.2547147.
46. Deepa B and Dr.Ramesh K. Preprocessed chb-mit scalp eeg database. 2021. doi: 10.21227/awcw-mn88. URL https://dx.doi.org/10.21227/awcw-mn88.
47. Palak Handa and Nidhi Goel. Peri-ictal and non-seizure eeg event detection using generated metadata. *Expert Systems*, page e12929. doi: https://doi.org/10.1111/exsy.12929. URL https://onlinelibrary.wiley.com/doi/abs/10.1111/exsy.12929.
48. Levin Kuhlmann, Philippa Karoly, Dean R Freestone, Benjamin H Brinkmann, Andriy Temko, Alexandre Barachant, Feng Li, Gilberto Titericz Jr, Brian W Lang, Daniel Lavery, et al. Epilepsyecosystem. org: crowd-sourcing reproducible seizure prediction with long-term human intracranial eeg. *Brain*, 141(9):2619–2630, 2018.
49. Spike-wave seizures and "proepileptic" activity in wag/rij rat model of absence epilepsy as recorded in the cortex, hippocampus and thalamus - mendeley data. https://data.mendeley.com/datasets/kdvxtc9xcg/. (Accessed on 04/14/2023).
50. Lucille A Lumley, Franco Rossetti, Marcio de Araujo Furtado, Brenda Marrero-Rosado, Caroline R Schultz, Mark K Schultz, Jerome Niquet, and Claude G Wasterlain. Dataset of eeg power integral, spontaneous recurrent seizure and behavioral responses following combination drug therapy in soman-exposed rats. *Data in brief*, 27:104629, 2019.
51. Petr Nejedly, Vaclav Kremen, Vladimir Sladky, Jan Cimbalnik, Petr Klimes, Filip Plesinger, Filip Mivalt, Vojtech Travnicek, Ivo Viscor, Martin Pail, et al. Multicenter intracranial eeg dataset for classification of graphoelements and artifactual signals. *Scientific data*, 7(1):1–7, 2020.
52. Mohammad Khubeb Siddiqui, Ruben Morales-Menendez, Xiaodi Huang, and Nasir Hussain. A review of epileptic seizure detection using machine learning classifiers. *Brain informatics*, 7(1):1–18, 2020.
53. Anna-Sophia Buschhoff, Igor Barg, Peer Wulff, and Andreas Bahr. Eeg of genetic absence epilepsy rats (gaers). 2020. doi: 10.21227/y1me-az56. URL https://dx.doi.org/10.21227/y1me-az56.
54. Khakon Das, Debashis Daschakladar, Partha Pratim Roy, Atri Chatterjee, and Shankar Prasad Saha. Epileptic seizure prediction by the detection of seizure waveform from the pre-ictal phase of eeg signal. *Biomedical Signal Processing and Control*, 57:101720, 2020.
55. Shivarudhrappa Raghu, Natarajan Sriraam, Yasin Temel, Shyam Vasudeva Rao, Alangar Satyaranjandas Hegde, and Pieter L Kubben. Performance evaluation of dwt based sigmoid entropy in time and frequency domains for automated detection of epileptic seizures using svm classifier. *Computers in biology and medicine*, 110:127–143, 2019.
56. Duanpo Wu, Zimeng Wang, Lurong Jiang, Fang Dong, Xunyi Wu, Shuang Wang, and Yao Ding. Automatic epileptic seizures joint detection algorithm based on improved multi-domain feature of ceeg and spike feature of aeeg. *IEEE Access*, 7:41551–41564, 2019.
57. Mustafa Talha Avcu, Zhuo Zhang, and Derrick Wei Shih Chan. Seizure detection using least eeg channels by deep convolutional neural network. In *ICASSP 2019-2019 IEEE international conference on acoustics, speech and signal processing (ICASSP)*, pages 1120–1124. IEEE, 2019.
58. Muhammad Bilal, Muhammad Rizwan, Sajid Saleem, Muhammad Murtaza Khan, Mohammed Saeed Alkatheir, and Mohammed Alqarni. Automatic seizure detection using multi-resolution dynamic mode decomposition. *IEEE Access*, 7:61180–61194, 2019.
59. Dhanalekshmi P Yedurkar and Shilpa P Metkar. Multiresolution approach for artifacts removal and localization of seizure onset zone in epileptic eeg signal. *Biomedical Signal Processing and Control*, 57:101794, 2020.
60. Peter Z Yan, Fei Wang, Nathaniel Kwok, Baxter B Allen, Sotirios Keros, and Zachary Grinspan. Automated spectrographic seizure detection using convolutional neural networks. *Seizure*, 71:124–131, 2019.
61. Gwangho Choi, Chulkyun Park, Junkyung Kim, Kyoungin Cho, Tae-Joon Kim, HwangSik Bae, Kyeongyuk Min, Ki-Young Jung, and Jongwha Chong. A novel multi-scale 3d cnn with deep neural network for epileptic seizure detection. In *2019 IEEE International Conference on Consumer Electronics (ICCE)*, pages 1–2. IEEE, 2019.
62. Jiuwen Cao, Jiahua Zhu, Wenbin Hu, and Anton Kummert. Epileptic signal classification with deep eeg features by stacked cnns. *IEEE Transactions on Cognitive and Developmental Systems*, 12(4):709–722, 2019.

63. Matthias Ihle, Hinnerk Feldwisch-Drentrup, César A Teixeira, Adrien Witon, Björn Schelter, Jens Timmer, and Andreas Schulze-Bonhage. Epilepsiae–a european epilepsy database. *Computer methods and programs in biomedicine*, 106 (3):127–138, 2012.
64. Björn Schelter, Matthias Winterhalder, Thomas Maiwald, Armin Brandt, Ariane Schad, Jens Timmer, and Andreas Schulze-Bonhage. Do false predictions of seizures depend on the state of vigilance? a report from two seizure-prediction methods and proposed remedies. *Epilepsia*, 47(12):2058–2070, 2006.
65. Piotr Zwoliński, Marcin Roszkowski, Jaroslaw Żygierewicz, Stefan Haufe, Guido Nolte, and Piotr J Durka. Open database of epileptic eeg with mri and postoperational assessment of foci—a real world verification for the eeg inverse solutions. *Neuroinformatics*, 8(4):285–299, 2010.